\documentclass{article}
\usepackage{arxiv}

\usepackage{hyperref}       
\usepackage{url}            
\usepackage{booktabs}       
\usepackage{amsfonts}       
\usepackage{nicefrac}       
\usepackage{microtype}      
\usepackage{lipsum}

\usepackage{fancyvrb}
\usepackage{xcolor}

\definecolor{mygreen}{HTML}{187418}

\usepackage{tikz}
\RequirePackage{calc}
\RequirePackage{verbatim}

\definecolor{shadecolor}{RGB}{240,240,240}
\definecolor{shadecolordark}{RGB}{200,200,200}
\definecolor{shadecolorlight}{RGB}{250,250,250}

\title{Starfish: A Prototype for Universal Preprocessing and
  Text-Embedded Programming}

\author{
  Vlado Ke\v{s}elj\thanks{\url{http://vlado.ca} ~or~ \url{http://vlado.cs.dal.ca}} \\
  Faculty of Computer Science\\
  Dalhousie University\\
  Halifax, NS B3H 4R2, Canada \\
  \texttt{vlado@dnlp.ca} or \texttt{vlado@cs.dal.ca} \\
}

\newcommand{\ltquestionmark}{{\tt <\tt ?}}

\begin{document}
\maketitle

\begin{abstract}
We present a novel concept of universal text preprocessing and
text-embedded programming (PTEP).  Preprocessing and text-embedded
programming has been widely used in programming languages and
frameworks in a fragmented and mutually isolated way.
The PTEP ideas can be found in the implementation of the
\TeX\ typesetting system; they are prominent in PHP and similar
web languages, and finally they are used in the Jupyter data science
framework.
This paper presents this area of research and related work in a more
unified framework, and we describe the implemented system Starfish
that satisfies the following novel principles of PTEP:
universality, update and replace modes, flexiblity, configurability,
and transparency.  We describe the operating model and design of
Starfish, which is an open-source system implementing universal
preprocessing and text-embedded programming in Perl.  The system is
transparent and its design allows direct implementation in other
programming languages as well.
\end{abstract}


\section{Introduction}

There is a wide need in computer science for {\em text Preprocessing}
and {\em Text-Embedded Programming (PTEP)}.
Transparency and ubiquity of
textual representation, similarly to the universality of natural
language, creates a widespread and uniform need to transform and
automatically manipulate text that we write before being processed in
some domain-specific way, such as programming language compilation,
document typesetting or visual rendering, and similar.

We define {\em text preprocessing} as any operation that takes text as
input and produces a similar text as output; and it serves as a way to
automate manual editing of text.  It is called ``preprocessing,''
since there is normally some standard use of that type of text that
would be called ``processing;'' such as compilation or interpretation
of a programming language, rendering of an HTML page, and similar.
A typical example of a preprocessor is the C programming language
preprocessor~\cite{c-preprocessor}, which is mostly used for simple
text inclusions or exclusions based on some configuration parameters
and simple text replacements, before the C program code is passed to
the compiler.

{\em Text-embedded programming} is a related but generally different
concept than text processing.  We define {\em text-embedded
programming} as any form of computer programming where code is
embedded in an arbitrary text, and can be executed in-place, in that
text context.
One of the first examples of text-embedded programming can be
considered the \TeX\ typesetting system by Donald Knuth, released in
1978~\cite{texbook}.
\TeX\ has its own language of annotating text to prepare it for
typesetting and printing in form of papers, books, and similar
documents.
The \TeX\ language also includes a macro language for
text transformation in-place, before the final preparation of the
output pages.
This macro part of the language is a form of preprocessing, but also
text-embedded programming because it is Turing-complete, and one can
write a general-purpose program in this language.
The \TeX{}book~\cite{texbook} contains a famous example of a table of
prime numbers, generated in the \TeX\ macro programming language,
written by Knuth.

The second, more obvious example of text-embedded programming is the
PHP programming language~\cite{php}.
A PHP program file is usually an HTML file with the snippets of PHP
code inserted in the file.  The file is processed, or we could say
preprocessed, before being delivered to the HTML browser in such way
that the PHP snippets are replaced with their output produced using
the command {\tt echo}.  The snippets are delimited with the strings
{\tt \ltquestionmark{}php} and {\tt ?>}, or simply with
\ltquestionmark\ and {\tt ?>}.
This model is particularly convenient for fast development of web
apps, where we can start with a static (pure) HTML page and
incrementally replace pieces with PHP snippets that will produce
dynamic PHP-generated HTML text.
A similar approach was used in ASP (Active Server
Pages)~\cite{asp-wiki} engine or JSP (JavaServer
Pages)~\cite{jsp-wiki}, both of which use \verb,<%, and \verb,%>,
delimiters for code snippets.

The third example of text-embedded programming is the project
Jupyter~\cite{jupyter}, formed in 2015 by Fernando
P\'erez~\cite{jupyter-wiki}, which supports inclusion of Python,
Julia, or R programming language snippets in a file called Jupyter
Notebook.
Although this example of text-embedded programming is not transparent
in the sense that a Jupyter Notebook is not a plain text file, it is
still very close to a text file (it is in the JSON format), marked in
a language called Markdown, which gets translated into HTML, and it
allows inclusion of arbitrary code in Python (or other allowed
language), that can be executed.  The result of the execution
is shown in the notebook itself.  This is a novelty, compared to PHP
for example, that we will call the {\em update mode}, vs. the {\em
replace mode} used in PHP, ASP, JSP, and similar template languages.

\subsection{Research Goals of Universal PTEP}

The main motivation for {\em Universal PTEP} (Preprocessing and
Text-Embedded Programming) is an initiative for creating a prototype
of universal system that could be easily used and adapted for an
arbitrary text.  Beside some major examples, such as \TeX, PHP, and
Jupyter, as mentioned, which have dedicated languages or are actually
centered around text-embedded programming, it would be beneficial to
have a system ready-to-be-used in an arbitrary text language, or {\em
style}, as we will call it, such as Makefiles, procmail recipes, Xfig
files, shell scripts, and many other text styles.
The approach is also attended to be used for any programming
language, web languages (e.g., HTML, CSS), data formats (e.g., JSON),
and typesetting languages (e.g., \TeX, \LaTeX).
Some ideas of this prototype were implemented starting from 1998, with
the system Starfish being named and released in
2001~\cite{starfish-cpan}.  It has been further developed since then,
and as of now, I am not aware of a similar effort in this area.

The main features that are the goal of this proposal and the Starfish
system are:
\begin{itemize}
\item {\bf Universal Preprocessing and Text-Embedded Programming
  (PTEP):} This is the goal of creating a universal system that can be used
  for PTEP for various types of text files (e.g., HTML, LaTeX, Java,
  Makefile, etc.)
\item {\bf Update and Replace Modes:} Starfish supports two modes of
  operation: the {\em replace mode} --- similarly to PHP or C
  preprocessor, where the snippets are replaced with the snippet
  output and the complete output is saved in the output file or
  produced to the standard output; or the {\em update mode} --- similarly
  to Jupyter, where the snippet output is appended to the snippet in
  the updated source file.
\item {\bf Flexible PTEP:} Starfish is flexible, in
  the sense that we can modify the patterns that are used to detect
  active code (snippets) in text.  The basic pattern used to detect and execute
  code snippets, can be generalized to make active ``hooks'' from any
  string, pair of delimiters, or regular expression pattern.
  Starfish also provides flexibility in defining the way snippets are
  evaluated.
\item {\bf Configurable PTEP:} Starfish allows user-defined
  configuration per directory, and it uses directory hierarchy for a more
  wide hierarchical configuration specification.
\item {\bf Transparent PTEP:} Starfish provides transparency in
  the sense that when a file with Starfish snippets is
  processed, assuming it is a static file, we do not need Starfish any
  more to use it.  For example, an HTML file is still an HTML file
  viewable by a browser, a \LaTeX\ file is still a \LaTeX\ file
  processable by \LaTeX, and so on.  As a comparison of a less
  transparent approach, a Jupyter file is a special-format JSON file,
  which needs to be processed to produce HTML, \LaTeX, or other forms
  usable by a user.
\item {\bf Embedded Perl:} Even though the main principles of Starfish
  could be implemented in many languages, Perl is particularly
  convenient for both Starfish implementation, use in code snippets,
  and configuration.  As a comparison, \TeX\ uses its own language for
  PTEP and it is difficult to use since its paradigm and notation
  style are so different from the main main-stream languages.
  The C preprocessor works well for C, but in attempts to use it in
  other systems, like {\tt Imake} for the {\tt Makefiles}, it was not
  widely adopted due to its restrictive nature and a significantly
  different context than C programs.
  Using a known, general-purpose language for specifying preprocessing
  steps and in the text-embedded snippets has clear advantages, and
  Perl succinctness and expressiveness in working with strings makes
  it an excellent candidate.
\end{itemize}

After this introduction, we will introduce main terminology and some
background on PTEP and related work in Section~\ref{sec:rw}.
In Section~\ref{sec:java}, we will present a case study of a Java
preprocessor functionality achieved using Starfish system, which we use
to describe fundamental modes of operation of Starfish.
Section~\ref{sec:universal} describes how we achieve universal PTEP by
using modifiable parameters such as delimiters and line comments.
We use example of Makefile and HTML styles to illustrate these
different contexts.  Finally, we give a conclusion with some ideas
about vision of PTEP for future work in Section~\ref{sec:conclusion}.

\section{Background and Related Work}\label{sec:rw}

We will describe in this section some background information on
Preprocessing and Text-Embedded Programming (PTEP), and some existing
related work in this area.  The PTEP area does not exist as a
recognized coherent area, but there has been a lot of fragmented
related work within the context of different programming languages,
and applied areas of Computer Science, such as in the web systems
development (PHP, ASP, JSP), software development (C preprocessor,
make, imake), electronic publishing (\TeX, \LaTeX), and machine
learning and data science (Jupyter).

\paragraph{Text and text files:} We define {\em text} to be any
string of characters, generally including the new-line character, and
it will typically be saved in a file, which we call a {\em text
file.}  We will assume characters to come from the ASCII set, but they
may include extended ASCII (i.e., numerical values from 0 to 255), or
they may have UTF-8 encoding, so characters may be from the Unicode
set.  A text is usually created manually in a
plain-text editor, such as {\tt emacs} or {\tt vi} in Linux or other
Unix-like systems, or {\tt notepad} in Windows OS. If the text follows
certain formal rules (grammar), we will say that text is in certain
{\em style}.  Otherwise, if we do not recognize a particular formal
grammar of the text, we will say that text has a {\em default style}.
It could be, for example, a general natural language text, such as
English, or any kind of text that is not on our list of recognized
styles.  We will also talk about specific styles, such as
the C-program style, if the text is a program in the C programming
language, a Java style, an HTML style, \LaTeX\ style, and similar.

\paragraph{Text preprocessing:} We define {\em text preprocessing} to
be an operation that takes text as input and produces a similar text
as output, and it serves as a way to automate manual editing of the
text.  Again, this is not a very precise definition, an we will have
to rely on some of our common sense and experience in recognizing what
constitutes preprocessing.  The name ``preprocessing'' comes from the
idea that this operations does not change the main style of the text,
and it is done before any proper {\em processing} designed for this
style of text, such as compilation of a C program, rendering of an
HTML page, or translating a LaTeX text into a PDF document.
A typical representative preprocessor is the C programming language
preprocessor~\cite{c-preprocessor,c-preprocessor-wiki}.

The generic potential of the C preprocessor has been recognized and it
was used independently of the C compiler; for example, in the Imake
system~\cite{imake-oreilly,imake-wiki}. Darren Miller~\cite{filepp}
made available Filepp---a generic file preprocessor, following closely
the C preprocessor syntax, with a number of generalizations.  The
Filepp preprocessor is written in Perl with intention to be used on
general text files and support for the HTML files.

\subsection{Text-Embedded Programming}

We define {\em text-embedded programming} as a form of computer
programming where programming source code is embedded in text of
arbitrary style, and this code can be executed in place; i.e., in the
original embedded context.

One could argue that any programming source code is embedded, since
code is generally mixed with documentation comments, but there is a
significant conceptual difference in thinking about a text file as a
program with comments, rather than a text of arbitrary style, with
some code snippets inserted.  We also leave some freedom in how the
code snippets are executed, to what purpose, or how they interact with the
surrounding text.  We will see soon some typical usages for such
snippets.

In text-embedded programming, programming source code is embedded in
text as a sequence of continuous text segments.  These segments are
sometimes called {\em code snippets}, {\em active code}, {\em live
code}, or {\em embedded code}.
We call the text outside the segments the {\em outer text}.
The code snippets are usually easily recognizable by defined text
delimiters, but depending
on the rules that we use, any text can be recognized as a snippet.
This is why the name {\em active code} is very appropriate:  The TeX
system uses a labeling of all characters at run-time that can denote
any character to be an `active' character, and as such initiate
special processing after the system reads this character.  This
character is also sometimes called an {\em escape} character.
A similar generalized approach is adopted in Starfish, in which the
active code is recognized by {\em hooks,} which can be strings, pairs
of begin-end strings, or regular expressions.

One of the first examples of text-embedded programming was the
\TeX\ typesetting system developed by Knuth, released in
1978~\cite{texbook}.  The system processes text and prepares it for
typesetting pages for print, but in the process it recognizes
\TeX\ commands by detecting the {\em escape} backslash character
({\tt\char"5C}), i.e., an {\em active} character, which triggers
special execution behaviour based mostly on macro expansions.
This macro expansion model can be regarded as a computation model;
i.e., a model for code execution, but it is difficult to learn for
programming purposes as indicated by the author itself.  There were
approaches to developing {\TeX} preprocessors in other languages
such as Lisp, as published by Iwesaki in 2002~\cite{iwesaki}.

Text-Embedded Programming is particularly popular and useful in the
context of HTML documents.  The HTML language was after its design
mostly used for creation of static documents, viewable and browsable
by users, and a very natural way to make the documents dynamic through
programming is by inserting code snippets in HTML pages.
Before serving the page for viewing and browsing, the snippets are
executed and replaced with their textual output, and the resulting
page is sent to the browser.
This model is used in the very popular PHP
language~\cite{php}, and also in ASP (Active Server
Pages)~\cite{asp-wiki} and JSP (JavaServer Pages)~\cite{jsp-wiki}.

\begin{table}\caption{Escape Strings in Some Systems}
\centering
\begin{tabular}{lcc}\toprule
System & \multicolumn{2}{c}{Escape Strings} \\
\cmidrule(r){2-3}
       & Begin & End \\
       \midrule
PHP    & \ltquestionmark           & \verb,?>, \\
       & \ltquestionmark\verb,php, & \verb,?>, \\
\midrule
ASP    & \verb,<%,                 & \verb,%>,     \\
JSP    & \verb,<%,                 & \verb,%>,     \\
ePerl  & \ltquestionmark           & \verb,!>,    \\
Text::Template     & \verb,{,         & \verb,}, \\
Text::Oyster       & \ltquestionmark  & \verb,?>, \\
\bottomrule
\end{tabular}\quad
\begin{tabular}{lcc}\toprule
System & \multicolumn{2}{c}{Escape Strings} \\
\cmidrule(r){2-3}
       & Begin & End \\
\midrule
HTML::EP           & \verb,<ep-perl>, & \verb,</ep-perl>, \\
\midrule
Starfish (default) & \ltquestionmark & \verb,!>,\\
Starfish (HTML style)    & \verb,<!--,\ltquestionmark & \verb,!>-->, \\
Starfish (user defined)  & \ltquestionmark\verb,sfish, & \verb,!>,   \\
Starfish (user defined)  & \vdots & \vdots\\
\bottomrule
\end{tabular}
\label{tab:delimiters}
\end{table}

The code snippets are marked in text with starting and ending
delimiters, which are arbitrary small strings.  Other than simple
markers for snippets, we can think of them as escape sequences that
toggle on and off code processing.  Table~\ref{tab:delimiters} shows
the live code (snippet) delimiters in some systems.
For example, the string delimiters are ``\ltquestionmark'' and
``\verb.?>.'' or ``\ltquestionmark\verb.php.'' and ``\verb.?>.'' in PHP,
``\verb,<%,'' and ``\verb,%>,'' in ASP, and ``\ltquestionmark'' and
``\verb,!>,'' in ePerl.
For example, in PHP, we could prepare an HTML document such as:
\begin{quote}
{\tt
<html><head><title>PHP Test</title></head>\\
<body>\\
\textcolor{red}{<\kern0pt?php} \textcolor{blue}{echo '<p>Hello
  World</p>'; }\textcolor{red}{?>}
</body></html>
}
\end{quote}
where we show snippet delimiters in red, and the snippet itself in
blue color.
After processing with the PHP interpreter, the following output
would be produced:
\begin{quote}{\tt
<html><head><title>PHP Test</title></head>\\
<body>\\
\textcolor{mygreen}{<p>Hello World</p>}\\
</body></html>
}\end{quote}
where we show the generated output in the green color.
Embedding the code in this way is sometimes called {\em escaping} because
a starting delimiter, such as ``\ltquestionmark'' serves as an escape
sequence, triggering special processing of the snippet.
Another kind of escaping, referred to as the {\em advanced escaping} in
PHP is illustrated with the following example:
\begin{quote}{\tt
Good \textcolor{red}{\ltquestionmark php}
\textcolor{blue}{ if (\$hour < 12) \{ }\textcolor{red}{?>} Morning
\textcolor{red}{\ltquestionmark php}\textcolor{blue}{ \} else \{ }
\textcolor{blue}{?>} Evening \textcolor{blue}{\ltquestionmark php}
\textcolor{red}{\}} \textcolor{blue}{?>}
}\end{quote}
We will refer to this kind of escaping as {\em inverted escaping.}
Inverted escaping can be interpreted in the following way:
The complete input text is treated as code in which the plain text,
i.e., the non-code text or {\em outer text}, is embedded between
`\verb,?>,' and `\ltquestionmark php' delimiters and it is translated into an
`\verb,echo ",{\em string}\verb,";,' statement; and similarly, any
part of the form `\verb,?> plain text ,\ltquestionmark' is interpreted
as the statement:
\begin{quote}{\tt
echo " plain text ";}\end{quote}
An implicit delimiter `\verb,?>,' is assumed at the beginning of the
text and an implicit delimiter `\ltquestionmark\verb,php,' is assumed
at the end of text.
Although this type of escaping is relatively easy to implement, we do
not use inverted escaping in Starfish since its benefits are not very
clear.
On the other hand, inverted escaping does not follow the principle that
each snippet should be a well-defined block of code.
If we want large pieces of outer text to be conditionally included or
excluded, Perl offers many string delimiting options for large text
segments, such as \verb,q/.../, and \verb,<<'EOT',, which can be used
in place of inverted escaping.

\subsection{Perl-based Embedded Programming}

The universal PTEP proposal is programming language independent, and
the prototype is easily adaptable to work with any programming
language.  Any higher-level language that provides easy manipulation of
strings, automated memory management, and run-time evaluation of code
are very suitable for PTEP.  The Perl programming language is
particularly suitable due to it string-processing efficient and
expressive string-processing functionalities.
For this reason, there has always been many Perl implementations of
preprocessing and some forms of text-embedded programming, including
templating modules.  Even the core Perl language always included simple
template-based string generation with interpolated strings, in which
a string such as
\begin{quote}
  \verb,"The flight $flightid arrives at $time.",
\end{quote}
is evaluated to: 
\verb,"The flight AC806 arrives at 11:26.", if the variables
\verb,$flightid, and \verb,$time, have values:
\begin{quote}
\begin{verbatim}
  $flightid = "AC806";
  $time     = "11:26";
\end{verbatim}
\end{quote}

An early system for embedded Perl in a fashion similar to PHP was
ePerl (Embedded Perl Language) by Ralf S. Engelshall~\cite{eperl}.
The language ePerl was developed in the period from 1996 to~1998.
The system was a binary package based on modified code of the Perl
interpreter, and as such had a relatively large memory footprint, or
was a ``heavy-weight'' implementation as sometimes
called~\cite{eperl-h}.  This approach had some other disadvantages
such as need for recompilation for each platform, and not keeping with
the evolution of the Perl language, unless regularly maintained.
These issues are addressed by implementing the system as a Perl
module; i.e., a language extension in Perl terminology.
For example, David Ljung Madison developed an ``ePerl
hack''~\cite{eperl-h} which is a Perl script of some 1400 lines that
has functionality similar to ePerl.
In comparison to Starfish, in addition to the ``heavy-weight''
implementation, ePerl does not support the update mode and is not
designed for a universal PTEP.

Text::Template~\cite{text-template} by Mark Jason Dominus
is another Perl module with similar functionality.
It is a very popular module designed to ``expand template text with
embedded Perl'', created in 1995 or 1996 and maintained with
contributions by many users until now.
An interesting and probably independent similarity is that Starfish
uses \verb,$O, as the output variable, while \verb,$OUT, is used in
Text::Template.
The default embedded code delimiters in Text::Template are `\verb,{,'
and `\verb,},', with an additional condition that braces have to be
properly nested.  For example, `\verb,{{{"abc"}}},' is a valid snippet
with delimiters.  The module allows the user to change the default
delimiters to other alternative delimiters.  The philosophy of
Text::Template module has a lot of similarity with Starfish, however
the Text::Template module is primarily meant to be used in templating
style; which means that a template file is created as a more passive
object and it always requires a handling Perl script to generate
the output target file.  An additional difference is that the
Text::Template module does not support the update mode.  The use of
default delimiters creates issues with JavaScript code, although there
are workarounds.  The system is not applied to many text styles
other than plain text and HTML.

Another well-known Perl module HTML::Mason~\cite{html-mason}, authored
by Jonathan Swartz, Dave Rolsky, and Ken Williams, can also be seen as
an embedded Perl system.  It is a larger system with the major design
objective to be a high-performance, dynamic, web-site authoring
system.

A relatively minimalistic approach is used in development of the
module Text::Oyster~\cite{oyster} by Steve McKay in the period
2000--3.  The module  is template module for evaluating Perl embedded
in text between delimiters `\ltquestionmark' and `\verb,?>,'.

HTML::EP~\cite{html-ep} is another Perl module for embedding Perl into
HTML.  Its specific approach is that code delimiters are HTML-like
tags that start with `{\tt ep-}'.  For example, comments are delimited
by {\tt<ep-comment>} and {\tt </ep-comment>}, and active code is
delimited with tags {\tt<ep-perl>} and {\tt</ep-perl>}.  The last
value in the embedded code is the generated string.  The module is
meant to be used in a dynamic way over the Apache web server and the
use of Apache module {\tt mod-perl}, so the documentation gives a nice
overview of how to set up a Perl module that supports embedded
programming to run efficiently in this setting.
The set of tags is further extended, so
it includes {\tt<ep-email>} for generating emails from a web page,
{\tt <ep-database>} and {\tt<ep-query>} for working with a database,
{\tt <ep-list>} for generating HTML lists, then conditionals, and so
on.  It is an interesting idea that in text embedding like this we can
modify the language to be simpler in some situations than Perl, but it
is still not clear that it is justified to introduce all these new
constructs, when equivalent Perl code constructs exist.

Starfish is a lighter-weight system than eperl or Mason, but it is
more flexible and universal than Text::Template, the ePerl
hack, and HTML::EP.
Starfish covers a larger set of text styles than other systems,
provides other unique innovations, such as more flexibility
in defining active code detection patterns, per-directory
configuration, update mode, and full embedding when compared to the other
systems.
Under {\em full embedding} we refer to capability that all
functionality and customizability, such as adaptation of patters, can
be achieved with code inside the snippets embedded in a text file.

\subsection{PTEP in the Update Mode}

We can have different ways in which embedded code is executed
and how its output is used.  For example, even the concept of Literate
Programming~\cite{litprog-wiki} introduced by Knuth
in~1984~\cite{litprog} can be considered to be text-embedded programming,
although the code is only executed after it is automatically gathered
into the source files, and then compiled.

All the systems discussed in this section so far support the execution
model that we call the {\em replace mode} of execution.
In the {\em replace mode,} the code snippets are replaced with the
output of those snippets, and the file produced in this way is either
sent over internet to a browser to be viewed, or saved into a target
file.  The Starfish system was designed to support a new mode of
operation, called the {\em update mode,} in addition to the replace
mode, from its initial release in 2001~\cite{starfish-cpan}.
This mode was briefly described in a Perl Journal article in
2005~\cite{starfish-tpj}.  The main
property of the update mode is that rather than replacing the code
snippets with their output, the output is appended to snippets.  This
has several advantages: (1)~we are not required to setup a translation
process from source files to target files, which makes the project
management simpler;
(2)~it provides an easy inspection of the embedded code
and the output it produces, which is very convenient in prototyping,
for example; and (3)~it provides an easy way for the system
to be used as a preprocessor for text files of arbitrary style.
We will describe in more detail the update mode, but we mention it in
the related work section since a well-known system
Jupyter~\cite{jupyter,jupyter-wiki}, released in 2015, operates in the
update mode.  The Jupyter system works on files called Jupyter
Notebooks, which are JSON-type files with a mixture of plain text and
embedded Python code.  The execution of the notebook appends the
output of embedded code immediately to the code.  This is used to
create documents in which the code and results of code execution are
intermixed.  We would describe this as text-embedded programming with
the update mode, with a minor exception that Jupyter Notebook itself
is not in plan-text format but needs a viewer software to be presented
in that way.

After this summary of related work, we can introduce Starfish by
using a case of Java preprocessor functionality.

\section{Case Study: Java Preprocessor}\label{sec:java}

In this section, we will use the Java style example to describe
details of the Starfish model, and how Starfish can be directly used
in Java preprocessing.  As we mentioned before, a preprocessing
example is the C preprocessor, which is a useful and unique feature of
the C programming language.  It is a part of the C compiler package,
but it is a simple language in its own, which does simple text
manipulation before feeding it to the proper C compiler.  One use of
the preprocessor is inclusion or exclusion of parts of code depending on
values of some configuration variables.  It preprocesses C source
code as a general text, without a detailed use of C syntax or
semantics.  It is sometimes criticized for not using deeper semantics
of the language, and it is also praised for the same reason because it
is very clear what it does and it can be used on text other than C
programs.
For example, it was used in the {\tt Imake}
system~\cite{imake-wiki,imake-oreilly} for preprocessing {\tt
Makefiles}~\cite{make-gnu,make-wiki}.  Java does not have a
preprocessor and it would be useful in some situations.

When developing code in Java, we may need two versions of the
source code: a test version to be used for testing and development, and a
release version to be the production release.  The test version could
carry around a lot of meta information on data structures, be able to
produce verbose debug code, make additional expensive run-time checks,
and similar, while the release version would be efficient and slim in
code size and running time.
This means that at various places in the source code, we need to write
two versions of code snippets: a test version and a release version,
and the appropriate version would be included everywhere based on the
value of some global variable.  This could be simulated using Java
constructs, but the release code would be unnecessarily bloated, and
running-time efficiency of the code would likely be negatively
affected.

As an example, let us consider the following simple Java code:
\begin{quote}
\begin{Verbatim}[commandchars=\\\{\}]
/**
   A simple Java file.
*/

public class simple \{

  public static in main(String[] args) \{

    \textcolor{red}{System.out.println("Test version");}
    \textcolor{blue}{System.out.println("Release version");}

    return 0;
  \}
\}
\end{Verbatim}
\end{quote}
where the red line would be included in the test version of the code,
and the blue line would be included in the release version of the
code.

One solution would be to use the C preprocessor.  However, the C
preprocessor is a part of the C compiler and it is not meant and not
convenient to use independently.  Its functionality is tailored to the
C language, and it is not as easy to use for general and more
flexible text processing that we may want to have.
We would argue that it is more convenient to write a new text
preprocessor from scratch in a language like Perl, than to rely on the
C preprocessor for this purpose.
That leads to the second solution: we could write an independent
preprocessor from scratch in text-friendly high-level language like
Perl.  An even more flexible idea is to have a general-purpose
preprocessing system.
The system {\tt m4}~\cite{m4-gnu,m4-wiki} is one such
system, but it is limited to general-purpose macro processing, has its
own, specific syntax, and it does not support the update mode of
operation.

\subsection{Fully-Embedded Preprocessor}

One approach to our preprocessing task is to implement a program
similar to the C preprocessor, which would read our Java source files
and produce other files to be used for compilation.  To distinguish
these two files, we would have the original, {\em meta-source} file,
and produced {\em target source} Java file.  One issue with this
approach is that we must now manage two files for each Java source
file, and the second issue is that this preprocessor would be one-off
program with its own syntax, and a more general solution would be
applicable to a wider set of situations.  We could emulate the
functionality of the C preprocessor, but designing a new universal
preprocessor would allow us to think bigger and aim at a more
open-ended general functionality.  Both of the issues are addressed
with a {\em fully-embedded preprocessor}, which combines preprocessing
instructions and the preprocessing result in the same file, and allows
for a quite general Perl preprocessing code.

The Starfish system provides this functionality.  Our example Java
file could be written in the following way using the Starfish
conventions:
\begin{quote}
\begin{Verbatim}[commandchars=\\\{\}]
/**
   A simple Java file.
*/
// Uncomment version:
\textcolor{blue}{//\ltquestionmark\   $Version = 'Test';    !>}
\textcolor{red}{//\ltquestionmark\ # $Version = 'Release'; !>}

public class simple \{

  public static int main(String[] args) \{

    \textcolor{mygreen}{//\ltquestionmark\ $O = "    ".($Version eq 'Test' ?}
    \textcolor{mygreen}{// 'System.out.println("Test version");' :}
    \textcolor{mygreen}{// 'System.out.println("Release version");' );}
    \textcolor{mygreen}{//!>}

    return 0;
  \}
\}
\end{Verbatim}
\end{quote}
Starfish code is embedded Perl code found between delimiters
\ltquestionmark\ and {\tt !>}, and it is commented out using the Java
line comment notation {\tt //}.  The blue and red lines are used to
choose version of the software that we want to produce.  The red line
contains code commented out in Perl, so that chosen version is the
``Test'' version.  The green snippet code shows how we can select the
appropriate line of Java and produce it.  The Perl variable \verb,$O,
is used as a special variable to specify the generated code.  Starfish
has also a command {\tt echo} that effectively appends to this
variable.

If the name of the Java source file is {\tt simple.java} then we can
process it in Starfish by running the command (\verb,$, is a shell
prompt):
\begin{verbatim}
 $ starfish simple.java
\end{verbatim}
As the result of preprocessing, we do not create a new file, but the
source file is updated.  This is what we call the {\em update mode},
which is the default mode of Starfish operation.  The reason why we
call Starfish a {\em fully-embedded preprocessor} is that all necessary
preprocessing code, including even customization of delimiters and
snippet evaluations, can be done within the snippets themselves.
After running the above command, the contents of the file {\tt
simple.java} is now:
\begin{quote}
\begin{Verbatim}[commandchars=\\\{\}]
/**
   A simple Java file.
*/
// Uncomment version:
\textcolor{blue}{//\ltquestionmark\   $Version = 'Test';    !>}
\textcolor{red}{//\ltquestionmark\ # $Version = 'Release'; !>}

public class simple \{

  public static int main(String[] args) \{

    \textcolor{mygreen}{//\ltquestionmark\ $O = "    ".($Version eq 'Test' ?}
    \textcolor{mygreen}{// 'System.out.println("Test version");' :}
    \textcolor{mygreen}{// 'System.out.println("Release version");' );}
    \textcolor{mygreen}{//!>}\textcolor{magenta}{//+}
    \textcolor{magenta}{System.out.println("Test version");//-}

    return 0;
  \}
\}
\end{Verbatim}
\end{quote}
We can see that the desired line of code has been generated and
inserted in the file (magenta-colored part).  The generated part is
delimited with strings {\tt//+} and {\tt//-}, so if we run the {\tt
starfish} again on the file, the file will not be changed because the
generated part would be replaced with the same generated string.
If by coincidence our output code contains the string {\tt//-},
Starfish will insert a number in the delimiters; e.g., {\tt//3+} and
{\tt//3-}, so that the ending delimiter does note conflict with
accidental match in the generated code.

If we comment out the `Test' line and uncomment the `Release' line in
the new {\tt simple.java} file as follows:
\begin{quote}
\begin{Verbatim}[commandchars=\\\{\}]
/**
   A simple Java file.
*/
// Uncomment version:
\textcolor{blue}{//\ltquestionmark\ # $Version = 'Test';    !>}
\textcolor{red}{//\ltquestionmark\   $Version = 'Release'; !>}

public class simple \{

  public static int main(String[] args) \{

    \textcolor{mygreen}{//\ltquestionmark\ $O = "    ".($Version eq 'Test' ?}
    \textcolor{mygreen}{// 'System.out.println("Test version");' :}
    \textcolor{mygreen}{// 'System.out.println("Release version");' );}
    \textcolor{mygreen}{//!>}\textcolor{magenta}{//+}
    \textcolor{magenta}{System.out.println("Test version");//-}

    return 0;
  \}
\}
\end{Verbatim}
\end{quote}
and run:
\begin{verbatim}
 $ starfish simple.java
\end{verbatim}
again, the file {\tt simple.java} file will look as follows:
\begin{quote}
\begin{Verbatim}[commandchars=\\\{\}]
/**
   A simple Java file.
*/
// Uncomment version:
\textcolor{blue}{//\ltquestionmark\ # $Version = 'Test';    !>}
\textcolor{red}{//\ltquestionmark\   $Version = 'Release'; !>}

public class simple \{

  public static int main(String[] args) \{

    \textcolor{mygreen}{//\ltquestionmark\ $O = "    ".($Version eq 'Test' ?}
    \textcolor{mygreen}{// 'System.out.println("Test version");' :}
    \textcolor{mygreen}{// 'System.out.println("Release version");' );}
    \textcolor{mygreen}{//!>}\textcolor{magenta}{//+}
    \textcolor{magenta}{System.out.println("Release version");//-}

    return 0;
  \}
\}
\end{Verbatim}
\end{quote}

Since we can include arbitrary Perl code in the snippets, including
imports of external libraries and code, this framework provides a very
general way of code preprocessing.  Starfish includes a few more
features to support wider management of code base within a directory,
which we will discuss in the next two subsections.

\subsection{Preprocessing Multiple Files}

If we want to preprocess a number of Java files in a project, it would
be tedious and error-prone to modify each of them to set them to the
appropriate Test or Release version.  There are several ways how this
problem could be solved and we will describe three of them:
\\(1) using Perl {\tt require} command,
\\(2) using Make and Starfish {\tt-e} option, and
\\(3) using the Starfish {\tt starfish.conf} configuration file.

\paragraph{(1) Using Perl {\tt require} command:}
We can have one {\tt\$Version} parameter controlling many files by
simply having a Perl file called {\tt configuration.pl} with the
following content:
\begin{quote}
\begin{verbatim}
#!/usr/bin/perl
$Version = 'Test'; # Test or Release
1;
\end{verbatim}
\end{quote}
and one of the first lines in each Java source file would be:
\begin{quote}
\begin{Verbatim}[commandchars=\\\{\}]
//\ltquestionmark\ require 'configuration.pl' !>
\end{Verbatim}
\end{quote}
In this way, we would have one point of control for the Test or
Release version of all files.

\paragraph{(2) Using Make and the Starfish {\tt -e} option:}
Starfish has the option {\tt -e} for an initial Perl code execution,
somewhat similar to Perl, and we can use it to set the Version
variable.  For example, if we use a Makefile to compile all Java files
in a project, we could add a preprocessing command for each of them in
the following way in the Makefile:
\begin{quote}
\begin{verbatim}
VERSION=Test
#VERSION=Release

simple.class: simple.java
        starfish -e='$$Version="$VERSION"' $<
        javac $<
\end{verbatim}
\end{quote}
We would again have one point of version control, this time in the
Makefile.

\paragraph{(3) Using the Starfish {\tt starfish.conf} configuration
file:}  The idea of using a Perl configuration file, as shown in~(1),
is so common in many situations that we use a standard name for the
configuration file called {\tt starfish.conf} to include this
information.  Similarly to (1), the contents of the file {\tt
starfish.conf} would be:
\begin{quote}
\begin{verbatim}
#!/usr/bin/perl
$Version = 'Test'; # Test or Release
1;
\end{verbatim}
\end{quote}
and one of the first lines in each Java source file would be:
\begin{quote}
\begin{verbatim}
//<? read_starfish_conf !>
\end{verbatim}
\end{quote}
This is the common way to represent per-directory configuration in
Starfish.  One important difference between this approach and the
earlier approach with the standard Perl configuration file (1) is that
{\tt read\_starfish\_conf} behaves in a special way.  Namely, the
command {\tt read\_starfish\_conf} will look for a file named {\tt
starfish.conf} in the current directory; if found, it will then look
for the same named file in the parent directory.  Again, if it is
found, it will look into the parent of the parent directory and so on
until it cannot find a file with that name, or until it reaches the
top directory in the file system.
After that, it will execute, or more precisely {\tt
require} in the Perl terminology, all found files {\tt starfish.conf}
from top to bottom.  Each file is executed in its own directory as the
current directory.  This provides for a hierarchical per-directory
configuration, with natural process of parameter inheritance and
override option in sub-directories.
A similar process is used sometimes in the system of
Makefile in a project with multiple directories~\cite{make-recursive},
and in the Imake system for Makefile
generation.\cite{imake-oreilly,imake-wiki}

\subsection{Replace Mode}

Finally, if we want to produce a version of Java code without
preprocessing code, we can use the Starfish {\it replace mode.}  In
this mode, the preprocessing code is removed as well as markup around
the generated code.  We must specify an output file in the replace
mode because we normally do not want to permanently loose the
preprocessing code.  For example, if we run the following command:
\begin{verbatim}
 $ starfish -replace -o=release/simple.java simple.java
\end{verbatim}
on the above file in which {\tt\$Version} variable is set to the value
{\tt "Release"}, the resulting file {\tt release/simple.java} would
contain the following contents:
\begin{quote}
\begin{verbatim}
/**
   A simple Java file.
*/
// Uncomment version:



public class simple {

  public static int main(String[] args) {

        System.out.println("Release version");

    return 0;
  }
}
\end{verbatim}
\end{quote}

With this, we conclude this Java preprocessor case study of general
PTEP as implemented in the Starfish system.

\section{Universal PTEP: Modifying Active Patterns and Evaluation}
\label{sec:universal}

\begin{figure}\centering
\begin{tikzpicture}
\newlength{\mycodewidth}
\setlength{\mycodewidth}{.6\linewidth}    
\node (code) [fill=shadecolorlight,draw=shadecolordark,inner sep=6pt]
{\begin{minipage}{\mycodewidth}
\begin{Verbatim}[commandchars=\\\{\}]
/**
   A simple Java file.
*/
// Uncomment version:
\textcolor{blue}{//\ltquestionmark}\textcolor{red}{\ \# \$Version = 'Test';    }\textcolor{blue}{!>}
\textcolor{blue}{//\ltquestionmark}\textcolor{red}{\   \$Version = 'Release'; }\textcolor{blue}{!>}

public class simple \{

  public static int main(String[] args) \{

    \textcolor{blue}{//\ltquestionmark}\textcolor{red}{\ \$O = "    ".($Version eq 'Test' ?}
    \textcolor{mygreen}{//}\textcolor{red}{ 'System.out.println("Test version");' :}
    \textcolor{mygreen}{//}\textcolor{red}{ 'System.out.println("Release version");' );}
    \textcolor{mygreen}{//}\textcolor{blue}{!>}\textcolor{magenta}{//+}
    \textcolor{violet}{System.out.println("Test version");}\textcolor{magenta}{//-}

    return 0;
  \}
\}
\end{Verbatim}
\end{minipage}};
\newlength{\x}\newlength{\y}\newlength{\dx}\newlength{\dy}
\path [fill=shadecolor,draw=shadecolordark] (code.north west)
 +(180:2pt) rectangle ([yshift=0.2pt]code.south west);
\path (code.north west) +(-2.2pt,-0.1pt) node [anchor=south west,
  fill=shadecolor,draw=shadecolordark,inner sep=3pt]
  {\parbox{\mycodewidth+8.4pt}{\ttfamily simple.java}};
\path [draw=blue] (code.north west)
  +(0.15cm,-2.05cm) rectangle +(0.15cm+8.5mm,-2.05cm+3.5mm);
\setlength{\x}{0.15cm+8.55mm}\setlength{\y}{-2.05cm}
\setlength{\dx}{45.3mm}\setlength{\dy}{3.5mm}
\path [draw=red] (code.north west) +(\x,\y) rectangle +(\x+\dx,\y+\dy);
\setlength{\y}{-2.47cm}
\path [draw=red] (code.north west) +(\x,\y) rectangle +(\x+\dx,\y+\dy);
\path [draw=blue] (code.north west)
  +(0.15cm,-2.47cm) rectangle +(0.15cm+8.5mm,-2.47cm+3.5mm);
\setlength{\x}{5.55cm}\setlength{\y}{-2.05cm}
\setlength{\dx}{4.2mm}\setlength{\dy}{3.5mm}
\path [draw=blue] (code.north west) +(\x,\y) rectangle +(\x+\dx,\y+\dy);
\setlength{\x}{5.55cm}\setlength{\y}{-2.47cm}
\path [draw=blue] (code.north west) +(\x,\y) rectangle +(\x+\dx,\y+\dy);
\setlength{\x}{0.9cm}\setlength{\y}{-4.75cm}
\setlength{\dx}{8.5mm}\setlength{\dy}{3.5mm}
\path [draw=blue] (code.north west) +(\x,\y) rectangle +(\x+\dx,\y+\dy);
\newcommand{\myr}[5]{\setlength{\x}{#2}\setlength{\y}{#3}
\setlength{\dx}{#4}\setlength{\dy}{#5}
\path [draw=#1] (code.north west) +(\x,\y) rectangle +(\x+\dx,\y+\dy);}
\myr{blue}{0.9cm}{-4.75cm}{8.5mm}{3.5mm}
\myr{blue}{1.35cm}{-5.9cm}{3.4mm}{3.5mm}
\myr{red}{1.8cm}{-4.75cm}{65mm}{3.5mm}
\myr{mygreen}{0.9cm}{-5.15cm}{4.4mm}{3.5mm}
\myr{red}{1.37cm}{-5.15cm}{76.0mm}{3.5mm}
\myr{mygreen}{0.9cm}{-5.52cm}{4.4mm}{3.5mm}
\myr{red}{1.37cm}{-5.52cm}{83.5mm}{3.5mm}
\myr{mygreen}{0.9cm}{-5.90cm}{4.4mm}{3.5mm}
\myr{magenta}{1.72cm}{-5.90cm}{7.4mm}{3.5mm}
\myr{violet}{0.9cm}{-6.30cm}{65.0mm}{3.5mm}
\myr{magenta}{7.42cm}{-6.30cm}{7.7mm}{3.5mm}
\draw[blue,->] (code.north west) ++(-0.3,-2cm)
 node[left](pref){Begin delimiters} -- +(10:.4);
\draw[blue,->] (pref.east) -- ++(-30:.45);
\draw[blue,->] (pref.east) -- ++(-64:2.6);
\draw[blue,->] (code.east) ++(-2,2)
 node[right](suff){End delimiters} -- +(177:2.3);
\draw[blue,->](suff.west) -- +(186:2.3);
\draw[blue,->](suff.west) -- +(207.5:7.6);
\draw[red,->] (code.north) ++(1,-1) node[above](snippet){Code
  Snippets} -- +(200:1.9);
\draw[red,->](snippet.south) -- +(220:1.7);
\draw[red,->](snippet.south) -- +(280:3.4);
\draw[mygreen,->] (code.north west) ++(-0.3,-5.3)
node[left](linec){Line comments} -- +(15:1.2);
\draw[mygreen,->] (linec.east) -- ++(-3:1.15);
\draw[mygreen,->] (linec.east) -- ++(-19:1.2);
\draw[magenta,->] (code.south east) ++(-1.3,2.1)
  node[right](outd){Output delimiters} -- +(178.5:6.55);
\draw[magenta,->] (outd.west) -- ++(192:0.85);
\draw[violet,->] (code.south) ++(1,0.8)
  node[right](output){Output} (output.north) -- +(130.5:0.8);
\end{tikzpicture}
\caption{Illustration of Starfish parameters in Java code with active
  Starfish code}\label{fig:java}
\end{figure}
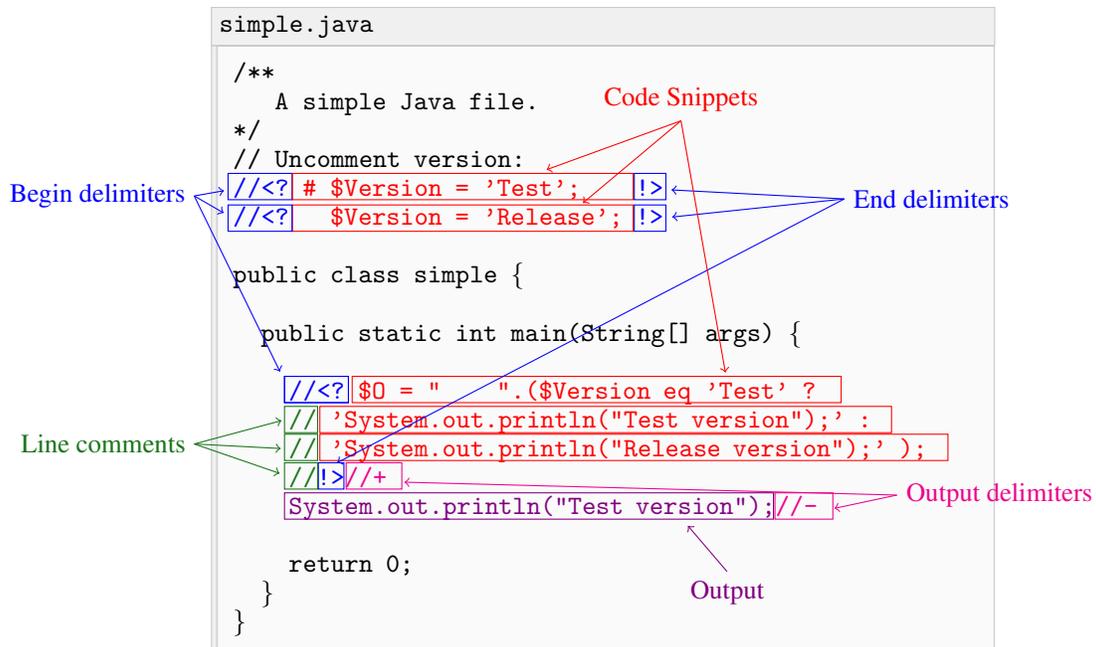

A universal PTEP feature of Starfish is implemented by allowing
run-time change of patterns for detecting active code and the model of
its evaluation.
For example, Figure~\ref{fig:java} shows active elements
of the Java code used in the previous section, with emphasized
strings used or generated by Starfish.
First, in order to detect active code (code snippets), Starfish relies
on the {\em begin} and {\em end delimiters} of the code snippets,
which we call {\em hooks}.  In Figure~\ref{fig:java}, the
begin-end hook consists of the strings
(`\verb,//,\ltquestionmark', `\verb,!>,').
In addition to this hook, there is another hook (`\ltquestionmark',
`\verb,!>,'), which is not used in this example.
Even though the first hook seems redundant, it is needed
for the replace mode, where we want the snippet output to replace the
line comment as well.
Starfish operates with a list of hooks, which can be modified at any
time.
In addition to the begin-end delimiter pairs, hooks can also be
expression hooks, which provide more flexibility.
Starfish is 
implemented in the object-oriented way, so during text processing
there is always a current Starfish-type object, accessible as the
variable \verb,$Star,, which contains the current state of parsing,
the list of active hooks, and other parameters.  The Starfish parser
always looks for the leftmost shortest match in the list of hooks in
order to locate the next code snippet.  Once the code snippet is
identified, the parser will also check for the optional immediately
following snippet output (only in the update mode), and remove it.

The output is detected using the {\em Output delimiters}.  Since the
output delimiters may include a generated number in order to avoid
conflict with the output, the output delimiters parameter
({\em OutDelimiters}) is actually an array of four strings:
(\verb,"//",, \verb,"+\n",, \verb,"//",, \verb,"-\n",), where the
\verb,\n, denotes the new-line character, and where
concatenation of the first two strings is the default starting output
delimiter, and the concatenation of the second two strings is the
ending output delimiter.  If there is a conflict with the snippet
output, a number is inserted between the starting and ending strings
to form non-conflicting delimiters.  The code snippet should be
commented out from the surrounding text, this is one reason why we need
information about the line comment (parameter {\em LineComment}).
Line comments are removed during the code preparation, the
code is evaluated, and the contents of the variable \verb,$O, is used
to determine the generated output.

All these parameters, hooks, output delimiters, line comment, as well
as the functions for code preparation and evaluation can be modified
to accommodate different text styles.  We can adjust these parameters
{\em ad hoc} in any snippet or configuration file to adapt to a new
text style.  A number of text styles (e.g., java, html, perl, python,
and makefile) are already provided in Starfish and we can set them by
calling the \verb,setStyle, function.  Otherwise, Starfish will
automatically choose the appropriate style based on the file
extension.  If the file extension is unknown, Starfish will set up the
{\em default} style.

Many text styles, such as Perl, Makefile, shell scripts, procmail
scripts, Python, and configuration files, use the hash symbol
(\verb,#,) as the line comment, so this line comment is chosen for the
default Starfish style.
Due to a specific syntax in Makefiles and Python, where the generated
output indentation is important, the output indentation is adjusted
according to the indentation of the lines in the code snippet.

\subsection{Default Style Example: Makefiles}

If a file name extension cannot be recognized, the Starfish will
process text in the default style, which relies on \verb,#, as the
line comment string, with the standard hooks and output delimiters.
{\tt Makefiles} are recognized as a separate style, but it is very
close to the default style so we will look a Makefile as an example.
The Makefiles are per-directory recipe files for the system~{\tt
make}~\cite{make-wiki,make-gnu}, which is used to run appropriate
commands to build files, including code compilation and linking,
testing, and other common tasks done using the command-line interface.
It can be used to efficiently compile C source files that need to be
compiled, \LaTeX\ files to be processed into PDF, figures produces etc.
Writing rules and dependencies for Makefiles can be tedious, and this
is why {\tt make} provides some help with its own advanced rules.
These advanced rules can be complex because they are a language of its
own, they are not always portable, and they do not provide flexibility
of a general programming language.  One attempt to address this
problem was the Imake system~\cite{imake-wiki}, which runs C
preprocessor on a Makefile before passing it to {\tt make}.
Some languages provide their own Makefile generators for
certain situations, such MakeMaker for Perl, but in a more general
sense we would like to have a Makefile preprocessor that would be
adaptable for handling many different projects, such as web site
deployment, or LaTeX typesetting.

As an example, let us say that we want to have a Makefile that would
adapt to any directory with a set of Java files, so that after running
\verb,make, all files that need to be compiled are compiled.
We could write a Makefile consisting only of the following Starfish
snippet:
\begin{quote}
\begin{Verbatim}[commandchars=\\\{\}]
#<? @javafiles = <*.java>;
#   echo "all: @javafiles\char"5Cn";
#   echo map \{ s/\char"5C.java$//; "$_.class: $_.java; javac $_.java\char"5Cn" \} @javafiles;
#!>
\end{Verbatim}
\end{quote}
The code above finds all \verb,.java, files in the current directory
and includes them into compilation.  After running:
\\\verb, $ starfish Makefile,
\\assuming that our directory contains files \verb,A.java,,
\verb,B.java, and \verb,C.java,, the Makefile is updated as follows:
\begin{quote}
\begin{Verbatim}[commandchars=\\\{\}]
#<? @javafiles = <*.java>;
#   echo "all: @javafiles\char"5Cn";
#   echo map \{ s/\char"5C.java$//; "$_.class: $_.java; javac $_.java\char"5Cn" \} @javafiles;
#!>#+
all: A.java B.java C.java
A.class: A.java; javac A.java
B.class: B.java; javac B.java
C.class: C.java; javac C.java
#-
\end{Verbatim}
\end{quote}

\subsection{HTML Style Example}

In our second example, we will show the HTML Starfish style.
The HTML style in Starfish (\verb,html,) is set up by setting the
begin and end hook delimiters to be `\verb,<!--<?,' and
`\verb,!>-->,', the output delimiters to be `\verb,<!-- + -->,' and
`\verb,<!-- - -->,', or more precisely as the four-tuple
(\verb|"<!-- +", " -->", "<!-- -", " -->"|),
and by not having line comments.  In this way, the Starfish code is
properly commented out and not visible in the browser in the update
mode of the HTML file.  The code would still work in the replace mode,
in which case the target HTML file would be generated by replacing
the code snippets with their output, in the same style as PHP, JSP,
and ASP.  Starfish can be run in the dynamic way using the Apache web
server, for example, where the output files would be produced
``on-the-fly'', again in the same fashion as PHP.

As an HTML example, let us consider the following short HTML file
with embedded Starfish code, which can be used to write a short blog:
\begin{quote}
\begin{verbatim}
<!--<? read_starfish_conf;
       $date_created = 'July 4, 2020';
       $title = 'My sample blog';
       echo blog_header(); !>-->

<p>This is an example blog post.  Below, you can find some source
code:
<!--<?
$_= <<'EOT';
/**
   A simple Java file with Test & Release
*/
// Uncomment version:
//<?   $Version = 'Test';    !>
//<? # $Version = 'Release'; !>

public class simple {

  public static int main(String[] args) {

    //<? $O = "    ".($Version eq 'Test' ?
    // 'System.out.println("Test version");' :
    // 'System.out.println("Release version");' );
    //!>
//etc...
EOT
echo "<PRE>".htmlquote($_)."</PRE>";
!>-->
\end{verbatim}
\end{quote}
We have a short code snippet at the beginning to generate start of the
HTML file using the function \verb,blog_header,, which is defined in
the file \verb,starfish.conf, so that we can use the same function for
all blog posts.  The function definition is as follows:
\begin{quote}
\begin{verbatim}
sub blog_header {
 return "<html><title>$title</title><body>\n".
        "Blog created: $date_created<br>\n".
        "Last update: ".file_modification_date()."\n".
        "<h1>$title</h1>\n";
}
\end{verbatim}
\end{quote}
The \verb,blog_header, function uses the Starfish-provided function
\verb,file_modification_date,, which supplies the last modification
date of the file.  It is interesting to note that Starfish itself
modifies the file; however, if there are no new update changes since
the last time we ran Starfish, new runs of Starfish will not
incorrectly update time to the time of the newest run.

We want to include Java source code in HTML.  This can be done using
the \verb,<PRE>, tag, however HTML still treats characters \verb,<,
and \verb,&, in a special way, so we are using Starfish-provided
function \verb,htmlquote, to escape those characters.

If we run Starfish on this file with:
\begin{verbatim}
  $ starfish blogexample.html
\end{verbatim}
the file will be updated as follows:
\begin{quote}
\begin{verbatim}
<!--<? read_starfish_conf;
$date_created = 'July 4, 2020';
$title = 'My sample blog';
echo blog_header(); !>--><!-- + --><html><title>My sample blog</title><body>
Blog created: July 4, 2020<br>
Last update: July 4, 2020
<h1>My sample blog</h1>
<!-- - -->

<p>This is an example blog post.  Below, you can find some source
code:
<!--<?
$_= <<'EOT';
/**
   A simple Java file with Test & Release
*/
// Uncomment version:
//<?   $Version = 'Test';    !>
//<? # $Version = 'Release'; !>

public class simple {

  public static int main(String[] args) {

    //<? $O = "    ".($Version eq 'Test' ?
    // 'System.out.println("Test version");' :
    // 'System.out.println("Release version");' );
    //!>
//etc...
EOT
echo "<PRE>".htmlquote($_)."</PRE>";
!>--><!-- + --><PRE>/**
   A simple Java file with Test &amp; Release
*/
// Uncomment version:
//&lt;?   $Version = 'Test';    !>
//&lt;? # $Version = 'Release'; !>

public class simple {

  public static int main(String[] args) {

    //&lt;? $O = &quot;    &quot;.($Version eq 'Test' ?
    // 'System.out.println(&quot;Test version&quot;);' :
    // 'System.out.println(&quot;Release version&quot;);' );
    //!>
//etc...
</PRE><!-- - -->
\end{verbatim}
\end{quote}
This is the output in the default Update mode.  The Replace mode would
produce a more clean HTML source without Starfish source, which could
be done for a production web site with the following command as an
example:
\begin{verbatim}
  $ starfish -replace -o=~/public_html/blog.html blogexample.html
\end{verbatim}

\section{Conclusion}\label{sec:conclusion}

In this paper, we want to present a proposal for a unified view of
{\em Preprocessing and Text-Embedded Programming (PTEP)} in different
contexts.  We give a background and a review of related work
showing that PTEP is widely used in different contexts.  Its use is
fragmented and the implementations use different and context-specific
syntax in most cases.
If we want to apply PTEP to a new context, we would always need to
implement a preprocessor from scratch.
We propose a universal approach to PTEP, with a number of novel
feature, and an open-source implementation of this approach named
Starfish~\cite{starfish,starfish-cpan,starfish-tpj}.

We discuss a case study of a Java preprocessor functionality achieved
with Starfish, which serves as an illustrative example.  After an
implementation description of Starfish, we show how it can be
customized for different text styles, and show additional examples of
the \verb,makefile, style, and \verb,html, style.  Starfish supports
additional styles, for which we do not have space here to discuss, but
the concept of universal PTEP opens many new future directions, such
as addressing further different text styles, use as a macro language,
and easy introduction of micro-languages for different local context.
In a grand view of things, universal PTEP is a step towards finding
commonality in some universal preprocessing and text-embedding task,
and defining them in a unified way in different contexts of textual
{\em computer instruction}, such as computer programming, typesetting,
configuration, and others.

\bibliographystyle{plain}
\bibliography{references}
\end{document}